\documentclass[%
 reprint,
superscriptaddress,
 amsmath,amssymb,
 aps,
]{revtex4-2}

\usepackage{graphicx}
\usepackage{dcolumn}
\usepackage{bm}
\usepackage{natbib}
\usepackage{float}

\makeatletter
\renewcommand*{\fnum@figure}{{\normalfont\bfseries \figurename~\thefigure}}
\makeatother

\begin{document}

\preprint{APS/123-QED}

\title{On-Chip Stimulated Brillouin Scattering via Surface Acoustic Waves}% Force line breaks with \\
\author{Govert Neijts}
\affiliation{Institute of Photonics and Optical Science (IPOS), School of Physics, The University of Sydney, NSW, 2006 Australia}
\affiliation{The University of Sydney Nano Institute (Sydney Nano), The University of Sydney, Camperdown, NSW, 2006 Australia}
\affiliation{Nonlinear Nanophotonics, MESA+ Institute of Nanotechnology, University of Twente, Enschede, The Netherlands}

\author{Choon Kong Lai}
\affiliation{Institute of Photonics and Optical Science (IPOS), School of Physics, The University of Sydney, NSW, 2006 Australia}
\affiliation{The University of Sydney Nano Institute (Sydney Nano), The University of Sydney, Camperdown, NSW, 2006 Australia}

\author{Maren Kramer Riseng}
\affiliation{Institute of Photonics and Optical Science (IPOS), School of Physics, The University of Sydney, NSW, 2006 Australia}
\affiliation{The University of Sydney Nano Institute (Sydney Nano), The University of Sydney, Camperdown, NSW, 2006 Australia}

\author{Duk-Yong Choi}
\affiliation{Laser Physics Centre, Research School of Physics, Australian National University, Canberra, ACT 2601, Australia}

\author{Kunlun Yan}
\affiliation{Laser Physics Centre, Research School of Physics, Australian National University, Canberra, ACT 2601, Australia}

\author{David Marpaung}
\affiliation{Nonlinear Nanophotonics, MESA+ Institute of Nanotechnology, University of Twente, Enschede, The Netherlands}

\author{Stephen J. Madden}
\affiliation{Laser Physics Centre, Research School of Physics, Australian National University, Canberra, ACT 2601, Australia}

\author{Benjamin J. Eggleton}
\affiliation{Institute of Photonics and Optical Science (IPOS), School of Physics, The University of Sydney, NSW, 2006 Australia}
\affiliation{The University of Sydney Nano Institute (Sydney Nano), The University of Sydney, Camperdown, NSW, 2006 Australia}

\author{Moritz Merklein}
\email{Electronic address: moritz.merklein@sydney.edu.au}
\affiliation{Institute of Photonics and Optical Science (IPOS), School of Physics, The University of Sydney, NSW, 2006 Australia}
\affiliation{The University of Sydney Nano Institute (Sydney Nano), The University of Sydney, Camperdown, NSW, 2006 Australia}

\begin{abstract}

Surface acoustic wave (SAW) devices are ubiquitously used for signal processing and filtering, as well as mechanical, chemical and biological sensing, and show promise as quantum transducers. However, nowadays most SAWs are excited and driven via electromechanical coupling and interdigital transducers (IDTs), limiting operation bandwidth and flexibility. Novel ways to coherently excite and detect SAWs all-optically interfaced with photonic integrated circuits are yet elusive. Backward Stimulated Brillouin scattering (SBS) provides strong coherent interactions between optical and acoustic waves in chip-scale waveguides, however, demonstrations have been limited to single longitudinal waves in the waveguide core.
Here, we numerically model and experimentally demonstrate surface acoustic wave stimulated Brillouin scattering (SAW-SBS) on a photonic chip. We designed and fabricated tailored waveguides made out of GeAsSe glass that show good overlap between SAWs at 3.81 GHz and guided optical modes, without requiring a top cladding. We measure a 225\,W$^{-1}$m$^{-1}$ Brillouin gain coefficient of the surface acoustic resonance and linewidth narrowing to 40\,MHz. Experimentally accessing this new regime of stimulated Brillouin scattering opens the door for novel on-chip sensing and signal processing applications, strong Brillouin interactions in materials that do not provide sufficient acoustic guidance in the waveguide core as well as excitation of surface acoustic waves in non-piezoelectric materials.
\end{abstract}

\maketitle

\section{\label{sec:1}Introduction}

Surface acoustic waves (SAWs) are harnessed in many signal processing and filtering applications for their high fidelity and compact footprint \cite{KenSAWTelecom, Ruppel2017a}, as well as a multitude of sensing applications - mechanical, structural, chemical, and biological - due to their sensitivity to the topography and surrounding of the surface \cite{MandalSAWsens, HaekwanStrainSAW,Huang2021}. The generation of SAWs is traditionally achieved in piezoelectric materials via interdigital transducers (IDTs). Exciting SAWs all-optically would enable new opportunities to achieve coherent coupling for quantum applications \cite{DumurSAWQuant, renningercoherentsawdev, renningerimpedancesawres}, more dynamic and tunable signal processing without bandwidth and frequency limitations set by the IDTs, and highly sensitive optical readout techniques for sensing applications. Furthermore, it has the potential to extend SAW technologies to non-piezo-electric materials.

The strongest nonlinear effect that couples acoustic and optical waves is known as stimulated Brillouin scattering (SBS) \cite{BoydNLO23}. Predicted more than 100 years ago \cite{Merklein2022}, it found many applications in fibre optical systems \cite{Kobyakov:10}, and more recently was successfully demonstrated in chip-scale waveguides \cite{EggletonBrillouinphotonics}. Inducing SBS on chip is challenging due to the short interaction length and the requirement of mode overlap between the optical and acoustic modes. The need for simultaneous guidance of the optical and acoustic mode, limits the Brillouin gain in most complementary metal-oxide semiconductor (CMOS) materials like SiN \cite{Gyger2019, Botter2022}, or requires under-etching in the case of silicon-on-insulator waveguides \cite{KuykenunderetchedSi, Kittlaus2015}. Soft chalcogenide glasses, on the contrary, provide simultaneous optical and acoustic guidance \cite{Pant2011, Poulton2013a}, resulting in demonstrations of as high as 50 dB on-chip Brillouin gain \cite{Choudhary50dB}. The large gain enabled record resolution Brillouin sensing of below a millimeter \cite{Zarifi2017}, however, only geometrical variations to the waveguide core could be sensed \cite{ZarifiBrSpec}.
\begin{figure*}
\includegraphics[width=\linewidth]{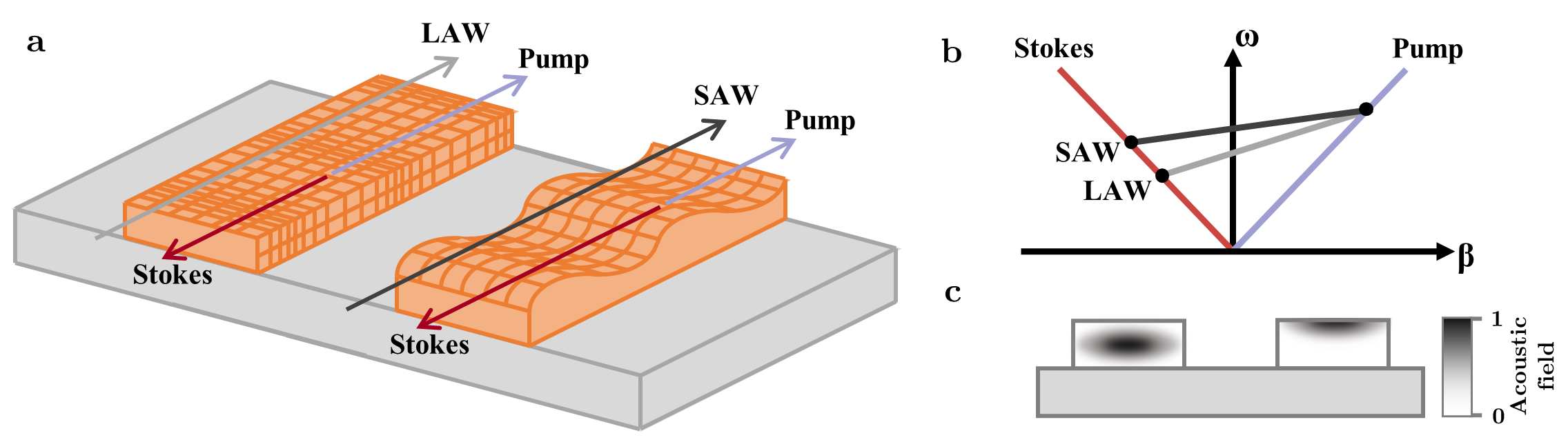}
\caption{\label{fig1} \textbf{Principle of Brillouin back-scattering by longitudinal and surface acoustic waves (a)} Schematic representation of on-chip Brillouin scattering by purely longitudinal acoustic waves (LAWs) (left) indicated as density fluctuations in the waveguide core and SAWs (right) travelling along the waveguide surface. \textbf{(b)} Dispersion diagram of the Brillouin scattering process from longitudinal and surface waves. \textbf{(c)} Illustration of the acoustic displacement profiles for longitudinal and surface waves in a waveguide.}
\end{figure*}
Hence, research on methods for achieving stimulated Brillouin interactions from acoustic waves propagating at the surface of the waveguide has drawn the attention of the scientific community. This new regime of Brillouin interactions based on SAWs on top of the waveguide would not only open new opportunities for high resolution sensing applications via coherent optical coupling but also overcome the material platform limitations caused by the lack of acoustic guidance while at the same time overcoming limitations imposed by traditional electromechanically excited SAWs. It was proposed theoretically that, for example, lithium niobate waveguides that do not show strong SBS gain would support strong SBS from SAWs \cite{Rodrigues:23}, and SAWs excited via IDTs in silicon-on-insulator waveguides enabled ultra-narrow bandwidth filter functions without the requirement of under-etching the silicon waveguide \cite{ZadokSAWSOI}. Weak spontaneous Brillouin light backscattering by surface acoustic waves has been studied in subwavelength optical fibre tapers \cite{BeugnotSAWBSfiber}, and forward Brillouin interaction have been observed at the surface of microspheres \cite{Bahl2011}; However, experimental demonstrations of optically excited and detected coherent SAWs via stimulated Brillouin scattering in a waveguide remain elusive. 

In this work, we present the first experimental observation of on-chip excitation and detection of surface acoustic waves via stimulated Brillouin scattering. This novel regime of Brillouin scattering was demonstrated in highly nonlinear Ge$_{11.5}$As$_{24}$Se$_{64.5}$ chalcogenide waveguides that were carefully designed and engineered to optimise the overlap with the guided optical mode and the acoustic wave propagating along the surface. Importantly, compared to many other chalcogenide materials such as As$_2$S$_3$ \cite{ChoiAs2S3, ChoiSU8} and As$_2$Se$_3$ \cite{Lai2023}, Ge$_{11.5}$As$_{24}$Se$_{64.5}$ does not require an over-cladding that might hamper the propagation of SAWs. We performed systematic numerical modeling of different waveguide geometries to enhance the overlap between optical and surface acoustic waves which match our experimental observation. Tailored waveguides were fabricated, and we measured 225\,W$^{-1}$m$^{-1}$ of SBS gain from surface waves in Ge$_{11.5}$As$_{24}$Se$_{64.5}$ waveguides with a cross-section of 2400\,nm by 116\,nm and an on-chip circuit length of 15 cm. Additionally, we studied the linewidth of the surface wave resonance and measured linewidth narrowing from 55\,MHz to 40\,MHz. Our first demonstration of stimulated surface wave Brillouin scattering marks a new regime for SBS and opens the door for many on-chip SAW-based applications.

\begin{figure*}
\includegraphics[width=\linewidth]{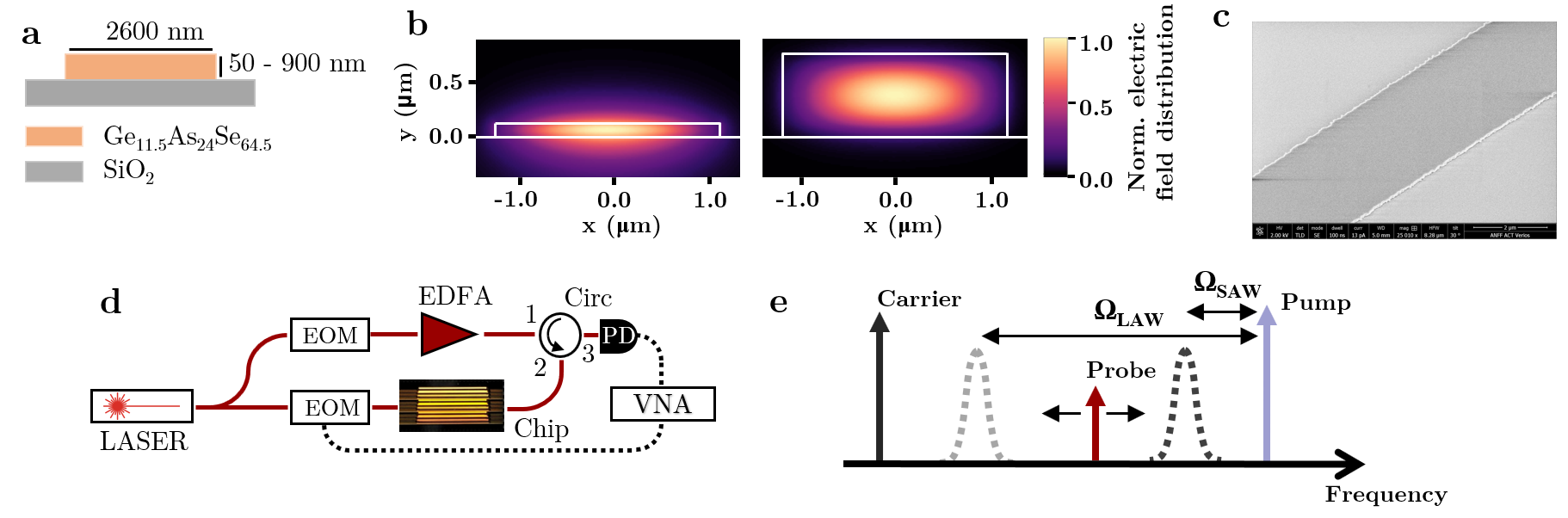}
\caption{\label{fig2} \textbf{Waveguide structure and schematic experimental setup. (a)} Schematic cross section of the investigated waveguide. \textbf{(b)} Fundamental optical mode profiles of thin (left) and thick (right) waveguide structure. \textbf{(c)} SEM image of typical fabricated waveguide. \textbf{(d)} Schematic setup for high-resolution pump-probe measurement. EOM, electro-optic modulator; EDFA, erbium-doped fibre amplifier; Circ, circulator; PD, photodetector; VNA, vector network analyser \textbf{(e)} Principle of the pump-probe measuring technique in the frequency domain.}
\end{figure*}
\section{\label{sec:2}Principle}
\subsection{Surface acoustic wave Brillouin scattering}

Brillouin scattering of light is a physical phenomenon that finds its origin in the third order nonlinearity of a material, and describes the coherent scattering of light from acoustic waves \cite{BoydNLO23}. It involves an incoming pump wave of frequency $\mathrm{\omega_p}$, an acoustic wave of frequency $\mathrm{\Omega_B}$, and a scattered optical Stokes wave that is Doppler shifted relative to the pump frequency $\mathrm{\omega_p}$ by the frequency of the acoustic wave $\mathrm{\omega_p - \Omega_B = \omega_s}$. The scattered Stokes light interferes with the incident light, reinforces the intensity of the scattering acoustic wave, that creates a positive feedback loop. This process is leading to  exponential amplification of the Stokes field, which is known as \textit{stimulated} Brillouin scattering . A main characteristic of SBS is the Brillouin frequency shift (BFS):

\begin{equation}
    \label{BrillouinFreq}
    \mathrm{\Omega_B=\frac{2n_{eff}v_q\omega_p}{c}},
\end{equation}

dependent on the effective refractive index of the waveguide $\mathrm{n_{eff}}$, the acoustic velocity $\mathrm{v_q}$ and the pump frequency $\mathrm{\omega_p}$, which depends on the material properties of the waveguide but is also sensitive to strain and temperature, making SBS a powerful sensing tool. Besides the BFS, the Brillouin gain coefficient $\mathrm{G_{SBS}}$, which describes the strength of the SBS interaction, and the gain linewidth of the Brillouin resonance $\Gamma_B$, determined by the phonon lifetime, are key when describing SBS in a waveguide. For backward SBS the optical pump wave and the Stokes wave typically couple to longitudinal acoustic waves (LAW) (Fig \ref{fig1}a, left), due to the large momentum difference of counter-propagating waves \cite{NLFO}, and the BFS is directly proportional to the longitudinal sound velocity in the medium.
Exciting their shear transverse counterparts, or more specifically coupling to surface acoustic waves (Fig. \ref{fig1}a, right) involves a smaller BFS (Fig. \ref{fig1}b), as the speed of surface waves is lower than the speed of longitudinal acoustic waves, typically between 0.87 and 0.95 times the shear speed of the material \cite{ViktorovRayleigh36}. That lower speed translates to a Brillouin frequency shift of the scattered Stokes light as can be seen in Eq.\ref{BrillouinFreq}. 

In the context of on-chip SBS, waveguides with geometries providing large overlap between the optical and the longitudinal acoustic field are used to enhance the Brillouin gain \cite{Poulton2013a}. These longitudinal acoustic waves can be localised in the core of the waveguide when the sound velocity in the waveguide core is smaller than in the surrounding material, and have a polarisation dominantly in the longitudinal direction (Fig. \ref{fig1}c). The displacement field of surface acoustic waves is, on the contrary, as the name suggests, localised at the surface of the core of the waveguide, and has a polarisation dominantly in the transverse direction. Enhancement of the Brillouin gain of surface waves requires optimisation of the waveguide geometry to improve the overlap between the optical and the surface acoustic field. Using highly nonlinear materials, such as chalcogenide glasses, have a large intrinsic SBS gain coefficient and hence are a promising candidate for demonstrating SAW-SBS \cite{PantOnchipsbs}.

\subsection{SAW-SBS waveguide design and numerical modelling}

The most common chalcogenide composition for SBS on-chip is As$_{2}$S$_{3}$ as it offers low propagation loss and large Brillouin gain \cite{Pant2011,Choudhary50dB}, and only recently on-chip SBS was demonstrated in other chalcognide glass compositions, such as GeSbS \cite{Song2021} and AsSe \cite{Lai2023}. The As$_{2}$S$_{3}$ platform requires a protective polymer layer to prevent chemical attack from the alkaline developer used in the photolithography process \cite{ChoiAs2S3, ChoiSU8}, which prevents As$_{2}$S$_{3}$ waveguides to be fabricated without any over-cladding which will suppress scattering from SAWs. So, although As$_{2}$S$_{3}$ is a perfect material for SBS based applications and showed record SBS gain, new waveguide fabrication methods, designs and materials are required for SAW-SBS. 
An alternative chalcogenide glass compositions, Ge$_{11.5}$As$_{24}$Se$_{64.5}$, poses a solution. Ge$_{11.5}$As$_{24}$Se$_{64.5}$ has a refractive index of 2.634 at infrared wavelengths, which is slightly higher than As$_{2}$S$_{3}$, similar elastic properties, and it does not require any protective layer during fabrication \cite{TWangGe}.\\
\begin{figure*}
\includegraphics[width=\linewidth]{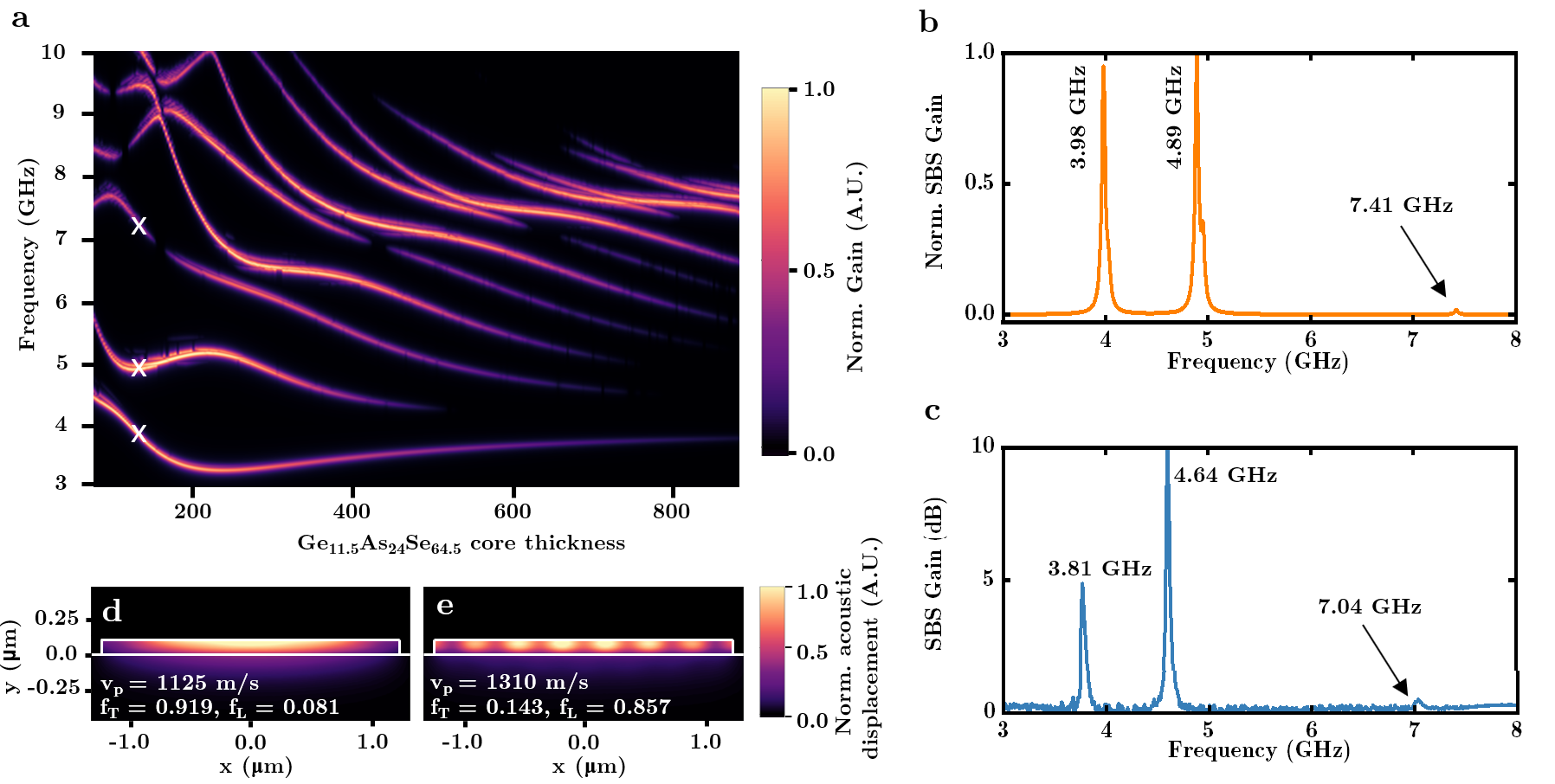}
\caption{\label{fig3}\textbf{Numerical and experimental findings of stimulated surface acoustic wave Brillouin scattering. (a)} Numerical simulations of Brillouin gain as function of frequency for the Ge$_{11.5}$As$_{24}$Se$_{64.5}$ core thickness varying from 70 to 900\,nm. The three white crosses illustrate the spectrum taken at 116\,nm waveguide thickness. \textbf{(b)} Simulated SBS spectrum for 116\,nm waveguide and \textbf{(c)} corresponding experimentally measured high-resolution SBS spectrum. \textbf{(d-e)} Acoustic displacement profile for the Brillouin resonance at 3.98\,GHz and 4.89\,GHz, respectively, with v$_p$, f$_T$ and f$_L$ the corresponding phase velocity, transverse and longitudinal polarisation fractions of the acoustic modes.}
\end{figure*}
With the aim of designing a waveguide structure that supports SAWs, numerical simulations of the Brillouin response are carried out in the open-source Numerical Brillouin Analysis Tool (NumBAT), a software tool developed for modelling of SBS in waveguides of arbitrary geometries \cite{SturmbergNumBAT}. NumBAT uses Finite Element Method (FEM) based mode solvers to obtain solutions for the optical and elastic modes of a given waveguide structure. It then evaluates overlap integrals of the (specified) optical and elastic modal fields that describe the interaction between the fields, in order to provide a spectrum of the SBS response of the waveguide. We simulated a rectangular waveguide structure comprising a 2600\,nm wide Ge$_{11.5}$As$_{24}$Se$_{64.5}$ core of different thicknesses on top of a SiO$_{2}$ substrate and importantly no over-cladding (Fig. \ref{fig2}a). All the relevant material parameters, refractive index ($n=2.634$), density ($\rho = 4495$ kg/m$^3$) and stiffness tensor components ($c_{11}=23.837$ GPa, $c_{12}=9.736$ GPa and $c_{44}=7.050$ GPa), have been taken from \cite{TWangGe}. Thick waveguide structures (Fig. \ref{fig2}b right) generally have a strong localisation of the fundamental optical mode in the core, whereas for thin structures (Fig. \ref{fig2}b left), the optical mode is less localised in the core and has a strong evanescent field, leading to potentially large overlap with surface acoustic modes that are localised at the surface of the waveguide. Hence, a sweep over the thickness of the waveguide core is performed, in order to find a thin structure that has high mode overlap between the optical mode and surface acoustic modes. A scanning electron microscope (SEM) image of a typical Ge$_{11.5}$As$_{24}$Se$_{64.5}$ waveguide is shown in Fig. \ref{fig2}c.

\subsection{Experimental setup}

In order to experimentally characterise and measure SBS in the simulated waveguide structures, we use a pump-probe setup (schematically outlined in Fig. \ref{fig2}d). The pump-probe method, first developed for characterising SBS in optical fibre \cite{Nikles1997}, is now the accepted standard for characterising Brillouin gain in integrated circuits \cite{BMORcomp1}. In essence it is a measurement of the beating between an optical carrier and an sweeping probe signal using a fast phototdetector and an electrical spectrum or network analyser. A schematic of the principle of the measurement technique is shown in Fig. \ref{fig2}d and e. Laser light (1550\,nm) that serves as a carrier, is split into two arms - a pump and a probe. The light in both arms is modulated using electro-optic modulators (EOM), filtered, and amplified by an erbium doped fibre amplifier (EDFA). On one side of the photonic chip, the strong optical pump wave is coupled to the waveguide via lensed tip fibres, in order to generate the Brillouin response of the material. From the other side of the chip a probe signal is coupled into the chip. A vector network analyser (VNA) is used to sweep the frequency of the probe sideband via an electro-optic modulator relative to the pump frequency, such that the difference between the pump and the probe is 1-10\,GHz. Using electrical spectrum or network analysers allows for high resolution measurement of the Brillouin spectrum. After passing through a circulator, an optical filter is used to remove any reflections from the optical pump wave and the beating between the probe carrier and the sweeping probe sideband is measured via a photodetector (PD). Changes in the amplitude of the probe signal due to Brillouin amplification are then measured by the VNA, providing a high resolution mapping of the Brillouin spectrum of the waveguide from the optical to the electrical domain

\begin{figure*}
\includegraphics[width=0.95\linewidth]{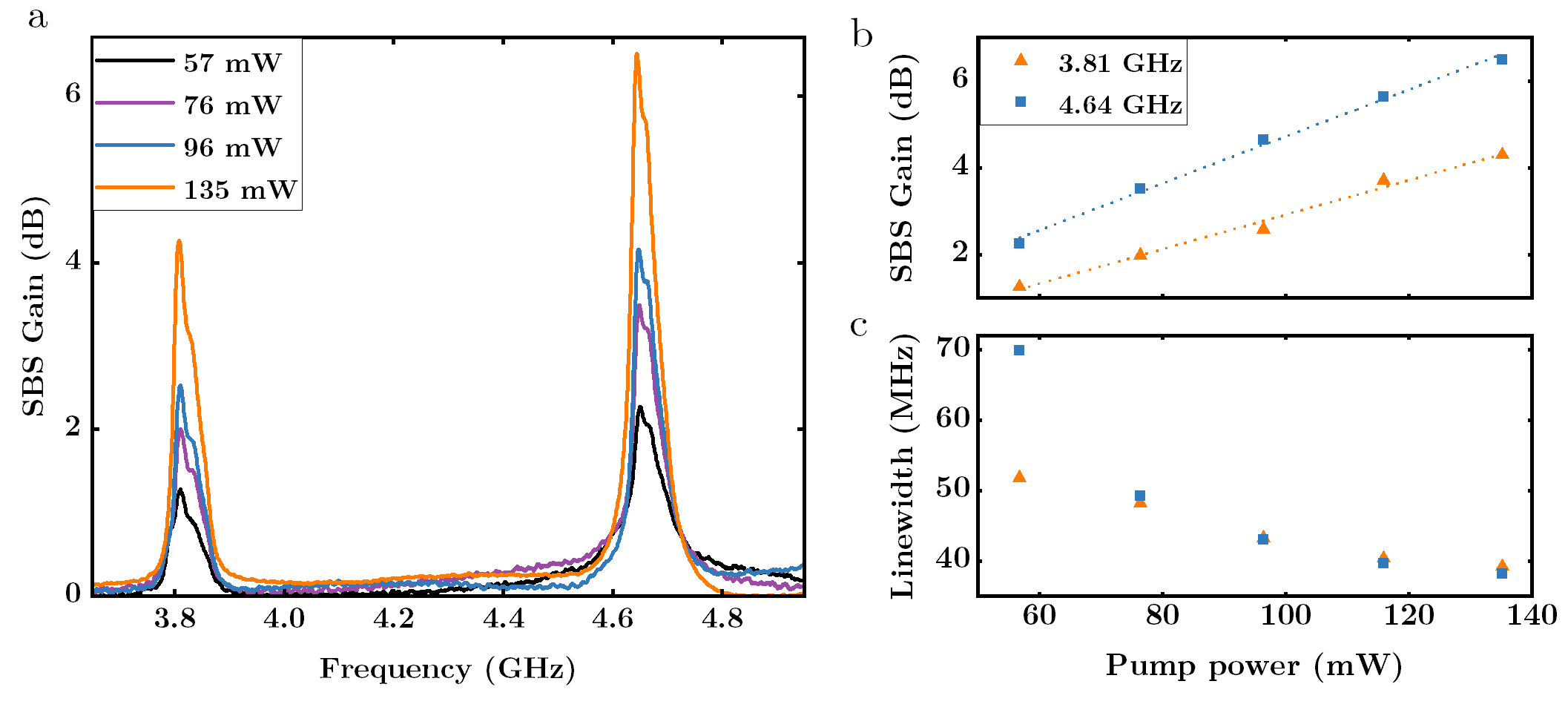}
\vspace*{-2mm}
\caption{\label{fig4}\textbf{Measurements for the Brillouin SAW gain coefficient and linewidth narrowing. (a)} Measurements of the SBS spectrum as a function of coupled pump power. \textbf{(b)} Peak gain values of the two modes for up to 135\,mW pump power and a linear fit. \textbf{(c)} Linewidth narrowing effect of the two modes for up to 135\,mW of pump power.}
\end{figure*}

\section{Results}
\subsection{Numerical study of Brillouin-active Ge$_{11.5}$As$_{24}$Se$_{64.5}$ waveguides}

Fig. \ref{fig3}a shows the simulated Brillouin gain as a function of frequency, for different Ge$_{11.5}$As$_{24}$Se$_{64.5}$ core thicknesses, ranging from 50\,nm - 900\,nm and a constant waveguide width of 2600\,nm. The simulation results show multiple Brillouin resonances appearing and disappearing as the waveguide thickness changes, and the appearance of lower frequency modes for thinner waveguide thicknesses. In order to optimise the structure for the support of the lower frequency surface acoustic modes, we select a thickness of 116\,nm, which has an effective refractive index $\mathrm{n_{eff}}$ of 1.641, of which the calculated Brillouin spectrum is presented in Fig. \ref{fig3}b. As Ge$_{11.5}$As$_{24}$Se$_{64.5}$ is not widely studied for Brillouin interactions \cite{KFrenchSBSGe}, required components of the photo-elastic tensor and acoustic loss tensor were not available in the literature. As these components solely determine the magnitude of the Brillouin gain, we normalise the simulation results presented in Fig. \ref{fig3}. The spectrum shows two strong resonances at 3.98\,GHz and 4.89\,GHz, and a weak resonance at 7.41 GHz. Further investigating the nature of the two strong modes, the acoustic displacement profiles (Fig. \ref{fig3}d and e) are calculated, including their acoustic velocity and polarisation fractions. From the mode profile in Fig. \ref{fig3}d, it is evident that the acoustic mode of the lowest frequency resonance has all the characteristics of a SAW. The acoustic mode profile is localised at the surface of the structure, the polarisation is dominantly in the transverse (y-) direction and the acoustic velocity is about 0.9 times the shear velocity of the material, verifying that the resonance is due to scattering from surface waves. The acoustic mode profile of the second peak has a transverse polarisation fraction that is reduced, and a longitudinal polarisation fraction that is increased, suggesting the resonance is a hybrid of shear transverse and longitudinal wave. This has been referred to as hybrid acoustic waves (HAWs), previously demonstrated in As$_{2}$S$_{3}$ \cite{NeijtsFiO}.

\subsection{\label{sec:3}SAW-SBS in Ge$_{11.5}$As$_{24}$Se$_{64.5}$ waveguides}
Based on the results from our numerical simulations, we fabricated a chip comprising a 116\,nm thick Ge$_{11.5}$As$_{24}$Se$_{64.5}$ layer, which has been deposited on top of an SiO$_{2}$ substrate, and etched to form a 2600\,nm wide waveguide. Initial measurements on the performance of the fabricated waveguides resulted in a 5\,dB coupling loss per facet and the propagation losses were estimated to be around 0.83 dB/cm, almost double previously reported values for the same material \cite{WangLosses1}, but can be attributed to the sidewall roughness observed in the SEM image (Fig. \ref{fig2}c.). 

High resolution pump-probe measurements of the SBS spectrum for a coupled pump and probe power of 150\,mW and 11\,mW, respectively, show two strong resonances around 4\,GHz, and a weak signal at 7.04\,GHz (Fig. \ref{fig3}c). The peak at 4.64\,GHz shows an SBS gain of 10 dB and has a linewidth of about 35\,MHz, whereas the peak at 3.81\,GHz has a gain of 5 dB and a linewidth of 40\,MHz. Comparing the experimental findings with our simulations, we find that our model matches well with the measured gain spectrum. The difference of about 0.2\,GHz between the experimental and numerical spectrum most likely finds its origin in uncertainty in the composition of the GeAsSe glass and the actual dimensions of the fabricated waveguide. The good agreement of simulations and measurements allows us to draw conclusion about the nature of the resonance at 3.81 GHz. Our simulations show that the mode at 3.81\,GHz has the typical characteristics of a SAW, demonstrating the first observation of on-chip excitation of surface acoustic waves by stimulated Brillouin scattering. Additionally, the same reasoning leads to determination of Brillouin scattering from hybrid shear longitudinal acoustic waves for the resonance at 4.64\,GHz.

We further investigated the characteristics of the observed SAW-SBS, in particular, the value of the Brillouin gain coefficient and the linewidth of the resonance. To that end, we measured the Brillouin spectrum for different pump powers using the pump-probe setup described in Fig. \ref{fig2}. The high resolution Brillouin spectra (Fig. \ref{fig4}a) show how the surface and hybrid acoustic resonances at 3.81 GHz and 4.64\,GHz, respectively, grow with an increasing coupled pump power. From the graph we obtain the peak gain values of the two resonances, and plot them as a function of pump power (Fig. \ref{fig4}b). Fitting the slope through the measured peak gain values, we find that both resonances grow linearly in a dB scale, or exponentially in a linear scale, verifying the nonlinear behaviour of the SBS process. From the slope we extract a Brillouin gain coefficient of $G_{SBS}=225$ W$^{-1}$m$^{-1}$ for the SAW mode and $G_{SBS}=305$ W$^{-1}$m$^{-1}$ for the HAW mode. By fitting Lorentzians to the measurements of the Brillouin responses, we extract the linewidth as a function of the pump power (Fig. \ref{fig4}c). Fig. \ref{fig4}c) shows narrowing of the linewidth of both peaks for increasing pump power, which is a typical characteristic of \textit{stimulated} Brillouin scattering \cite{GaetaBoydLinew} with both measured peaks ultimately narrowing to 40\,MHz.

\section{\label{sec:4}Conclusion and Discussion}
To summarise, in this work, we have demonstrated a first observation of on-chip stimulated Brillouin scattering by surface acoustic waves. Thin, 116\,nm waveguides made from nonlinear Ge$_{11.5}$As$_{24}$Se$_{64.5}$ chalcogenide glass enabled strong overlap between optical and acoustic waves propagating at the waveguide surface. We used Finite Element Method simulations, resulting in a comprehensive analysis of the Brillouin response of said waveguides, which were confirmed by experimental measurements. We observed surface acoustic wave stimulated Brillouin scattering, with a resonance frequency at 3.81 GHz. The SBS gain was measured for a range of pump powers, providing us with an extracted gain coefficient of $G_{SBS}=225$ W$^{-1}$m$^{-1}$ and a linewidth of 40 MHz. 
These initial results exemplify the remarkable potential embedded in the novel approach of surface acoustic wave excitation. Future steps towards enhancing the Brillouin gain for stronger interaction with surface waves involve optimising etching conditions for reduced sidewall roughness, and with that propagation losses, and general improvements to the waveguide structure in order to make the platform more robust. Additionally, modeling the environmental influence on the surface acoustic resonance of the Brillouin spectrum provides a pathway towards on-chip sensing experiments, manifesting as a first application of optically excited surface acoustic waves via stimulated Brillouin scattering.\\

\begin{acknowledgements}
We acknowledge the support of the Australian National Fabrication Facility (ANFF) OptoFab ACT Node in carrying out this research. 
\end{acknowledgements}

\section*{Funding Information}
This work has been supported through the European Research Council Consolidator Grant (101043229 TRIFFIC), Australian Research Council (DP220101431) and Office of Naval Research (N00014-23-1-2597).

\typeout{}
\bibliography{apssamp}

\end{document}